\documentclass[a4paper,12pt,notitlepage]{article}
\usepackage[utf8]{inputenc}
\usepackage[T1]{fontenc}
\usepackage[english]{babel}
\usepackage{amsmath}
\usepackage{booktabs}
\usepackage{xspace}
\usepackage{multirow}
\usepackage{tabularx}
\usepackage{upgreek}  
\usepackage{ifthen} 
\usepackage{graphicx}
\usepackage{mathrsfs}
\usepackage{lineno}
\usepackage{cite}
\usepackage{placeins}
\usepackage{amsmath}
\usepackage{subfig}

\newboolean{uprightparticles}
\setboolean{uprightparticles}{true}

\def\beq{\begin{equation}}
\def\eeq#1{\label{#1}\end{equation}}
\def\eeqn{\end{equation}}

\def\beqa{\begin{eqnarray}}
\def\eeqa#1{\label{#1}\end{eqnarray}}
\def\eeqan{\end{eqnarray}}

\let\bar=\overbar

\def\D{{\cal D}}

\def\Dslash{\not{\hbox{\kern-4pt $D$}}}
\def\dslash{\not{\hbox{\kern-2pt $\del$}}}

\def\BR{\mbox{\rm BR}}

\def\msb{{\bar{\ssstyle M \kern -1pt S}}}

\def\lhcb {LHCb\xspace}
\def\ux85 {UX85\xspace}

\def\lhc {LHC\xspace}

\def\cms {CMS\xspace}

\ifthenelse{\boolean{uprightparticles}}%
{

 \def\Pmu         {\ensuremath{\upmu}\xspace}

 \def\Ppi         {\ensuremath{\uppi}\xspace}

 \def\Ppsi        {\ensuremath{\uppsi}\xspace}

 \def\PDelta      {\ensuremath{\Delta}\xspace}                 
 \def\PXi      {\ensuremath{\Xi}\xspace}                 
 \def\PLambda      {\ensuremath{\Lambda}\xspace}                 
 \def\PSigma      {\ensuremath{\Sigma}\xspace}                 
 \def\POmega      {\ensuremath{\Omega}\xspace}                 
 \def\PUpsilon      {\ensuremath{\Upsilon}\xspace}                 
 
 \def\PB      {\ensuremath{\mathrm{B}}\xspace}                 
                  
 \def\PD      {\ensuremath{\mathrm{D}}\xspace}

 \def\PJ      {\ensuremath{\mathrm{J}}\xspace}                 
 \def\PK      {\ensuremath{\mathrm{K}}\xspace}                 
                  
 \def\PM      {\ensuremath{\mathrm{M}}\xspace}                 
 \def\PN      {\ensuremath{\mathrm{N}}\xspace}

 \def\PX      {\ensuremath{\mathrm{X}}\xspace}

 \def\Pb      {\ensuremath{\mathrm{b}}\xspace}                 
 \def\Pc      {\ensuremath{\mathrm{c}}\xspace}                 
                  
 \def\Pe      {\ensuremath{\mathrm{e}}\xspace}                 
 \def\Pf      {\ensuremath{\mathrm{f}}\xspace}

 \def\Pi      {\ensuremath{\mathrm{i}}\xspace}

 \def\Pp      {\ensuremath{\mathrm{p}}\xspace}

}
{

 \def\Pmu         {\ensuremath{\mu}\xspace}

 \def\Ppi         {\ensuremath{\pi}\xspace}

 \def\Ppsi        {\ensuremath{\psi}\xspace}                 
                  
 \mathchardef\PDelta="7101
 \mathchardef\PXi="7104
 \mathchardef\PLambda="7103
 \mathchardef\PSigma="7106
 \mathchardef\POmega="710A
 \mathchardef\PUpsilon="7107
                  
 \def\PB      {\ensuremath{B}\xspace}                 
                  
 \def\PD      {\ensuremath{D}\xspace}

 \def\PJ      {\ensuremath{J}\xspace}                 
 \def\PK      {\ensuremath{K}\xspace}                 
                  
 \def\PM      {\ensuremath{M}\xspace}                 
 \def\PN      {\ensuremath{N}\xspace}

 \def\PX      {\ensuremath{X}\xspace}

 \def\Pb      {\ensuremath{b}\xspace}                 
 \def\Pc      {\ensuremath{c}\xspace}                 
                  
 \def\Pe      {\ensuremath{e}\xspace}                 
 \def\Pf      {\ensuremath{f}\xspace}

 \def\Pi      {\ensuremath{i}\xspace}

 \def\Pp      {\ensuremath{p}\xspace}

}

\def\epem       {\ensuremath{\Pe^+\Pe^-}\xspace}

\def\mumu       {\ensuremath{\Pmu^+\Pmu^-}\xspace}

\def\bquark    {\ensuremath{\Pb}\xspace}
\def\bquarkbar {\ensuremath{\overline \bquark}\xspace}
\def\bbbar     {\ensuremath{\bquark\bquarkbar}\xspace}

\def\pion  {\ensuremath{\Ppi}\xspace}

\def\pip   {\ensuremath{\pion^+}\xspace}
\def\pim   {\ensuremath{\pion^-}\xspace}
\def\pipi  {\ensuremath{\pion^+\pion^-}\xspace}

\def\kaon  {\ensuremath{\PK}\xspace}
  \def\Kbar  {\kern 0.2em\overline{\kern -0.2em \PK}{}\xspace}

\def\Kz    {\ensuremath{\kaon^0}\xspace}
\def\Kzb   {\ensuremath{\Kbar^0}\xspace}
\def\KzKzb {\ensuremath{\Kz \kern -0.16em \Kzb}\xspace}
\def\Kp    {\ensuremath{\kaon^+}\xspace}
\def\Km    {\ensuremath{\kaon^-}\xspace}

\def\KpKm  {\ensuremath{\Kp \kern -0.16em \Km}\xspace}

  \def\Dbar    {\kern 0.2em\overline{\kern -0.2em \PD}{}\xspace}
\def\D       {\ensuremath{\PD}\xspace}
\def\Db      {\ensuremath{\Dbar}\xspace}
\def\Dz      {\ensuremath{\D^0}\xspace}
\def\Dzb     {\ensuremath{\Dbar^0}\xspace}
\def\DzDzb   {\ensuremath{\Dz {\kern -0.16em \Dzb}}\xspace}
\def\Dp      {\ensuremath{\D^+}\xspace}
\def\Dm      {\ensuremath{\D^-}\xspace}

\def\DpDm    {\ensuremath{\Dp {\kern -0.16em \Dm}}\xspace}

\def\B       {\ensuremath{\PB}\xspace}
  \def\Bbar    {\kern 0.18em\overline{\kern -0.18em \PB}{}\xspace}

\def\Bu      {\ensuremath{\B^+}\xspace}

\def\Bp      {\ensuremath{\Bu}\xspace}

\def\jpsi     {\ensuremath{{\PJ\mskip -3mu/\mskip -2mu\Ppsi\mskip 2mu}}\xspace}
\def\psitwos  {\ensuremath{\Ppsi(2S)}\xspace}

  \def\Y#1S{\ensuremath{\PUpsilon{(#1S)}}\xspace}

\def\proton      {\ensuremath{\Pp}\xspace}
\def\antiproton  {\ensuremath{\overline \proton}\xspace}

\def\Lbar {\ensuremath{\kern 0.1em\overline{\kern -0.1em\Lambda\kern -0.05em}\kern 0.05em{}}\xspace}

\def\BF         {{\ensuremath{\cal B}\xspace}}

\def\BR         {\BF}

\def\to                 {\ensuremath{\rightarrow}\xspace}

\def\AT#1     {\ensuremath{A_{\mathrm{T}}^{#1}}\xspace}           

\def\C#1      {\ensuremath{\mathcal{C}_{#1}}\xspace}                       
\def\Cp#1     {\ensuremath{\mathcal{C}_{#1}^{'}}\xspace}                    
\def\Ceff#1   {\ensuremath{\mathcal{C}_{#1}^{\mathrm{(eff)}}}\xspace}        
\def\Cpeff#1  {\ensuremath{\mathcal{C}_{#1}^{'\mathrm{(eff)}}}\xspace}       
\def\Ope#1    {\ensuremath{\mathcal{O}_{#1}}\xspace}                       
\def\Opep#1   {\ensuremath{\mathcal{O}_{#1}^{'}}\xspace}                    

\newcommand{\tev}{\ensuremath{\mathrm{\,Te\kern -0.1em V}}\xspace}
\newcommand{\gev}{\ensuremath{\mathrm{\,Ge\kern -0.1em V}}\xspace}
\newcommand{\mev}{\ensuremath{\mathrm{\,Me\kern -0.1em V}}\xspace}
\newcommand{\kev}{\ensuremath{\mathrm{\,ke\kern -0.1em V}}\xspace}
\newcommand{\ev}{\ensuremath{\mathrm{\,e\kern -0.1em V}}\xspace}
\newcommand{\gevc}{\ensuremath{{\mathrm{\,Ge\kern -0.1em V\!/}c}}\xspace}
\newcommand{\mevc}{\ensuremath{{\mathrm{\,Me\kern -0.1em V\!/}c}}\xspace}
\newcommand{\gevcc}{\ensuremath{{\mathrm{\,Ge\kern -0.1em V\!/}c^2}}\xspace}
\newcommand{\gevgevcccc}{\ensuremath{{\mathrm{\,Ge\kern -0.1em V^2\!/}c^4}}\xspace}
\newcommand{\mevcc}{\ensuremath{{\mathrm{\,Me\kern -0.1em V\!/}c^2}}\xspace}

\def\nb {\ensuremath{\rm \,nb}\xspace}

\def\invpb {\ensuremath{\mbox{\,pb}^{-1}}\xspace}

\newcommand{\stat}{\ensuremath{\mathrm{(stat)}}\xspace}
\newcommand{\syst}{\ensuremath{\mathrm{(syst)}}\xspace}

\def\gsim{{~\raise.15em\hbox{$>$}\kern-.85em
          \lower.35em\hbox{$\sim$}~}\xspace}
\def\lsim{{~\raise.15em\hbox{$<$}\kern-.85em
          \lower.35em\hbox{$\sim$}~}\xspace}

\def\sqs   {\ensuremath{\protect\sqrt{s}}\xspace}

\def\pt         {\mbox{$p_{\rm T}$}\xspace}

\newcommand{\lum} {\ensuremath{\mathcal{L}}\xspace}

\def\tell1  {TELL1\xspace}
\def\ukl1   {UKL1\xspace}

\def \Xa{\ensuremath{\PX(3872)}\xspace}
\def \Xb{\ensuremath{\PX(4140)}\xspace}
\def \Xc{\ensuremath{\PX(4274)}\xspace}
\def \BToKJpsiPhi{\ensuremath{\Bu \to \Kp \jpsi \phi}\xspace}
\def \BToXbK{\ensuremath{\Bu \to \Kp \Xb}\xspace}
\def \BToXcK{\ensuremath{\Bu \to \Kp \Xc}\xspace}
\def \BToXaK{\ensuremath{\Bu \to \Kp \Xa}\xspace}
\def \XbToJpsiPhi{\ensuremath{\Xb \to \jpsi \phi}\xspace}
\def \XcToJpsiPhi{\ensuremath{\Xc \to \jpsi \phi}\xspace}
\def \XaToJpsiPiPi{\ensuremath{\Xa \to \jpsi \pip\pim}\xspace}

\author{{\Large A. Augusto Alves Jr, on behalf of LHCb Collabration}\\
Universita degli studi di Roma ``La Sapienza'' and INFN sezione di Roma\\
P.le Aldo Moro, 2 - 00185 Roma - Italy\\
\vspace{1.5cm}\\
\textbf{Presented at 5th International Workshop on Charm Physics}\\
Honolulu, Hawaii, May 17 2012}

\title{\large New results on the spectroscopy of X, Y, Z states from LHC experiments\\}

\begin{document}

\maketitle

\begin{abstract}
The main results from LHC experiments on XYZ
charmonium-like candidates are summarized.
\end{abstract}
\clearpage

\section{Introduction}

According to our current understanding, the forces responsible to bind quarks into hadrons are
described by the non-Abelian field theory called Quantum Chromodynamics
(QCD). In QCD-motivated quark potential models, the quarkonia states are
described as a quark-antiquark pair bound by an interquark force with a short-
distance behavior that is approximately Coulombic, plus an increasing
confining potential that dominates at large separations. In one of the simplest approaches,
the energy levels can be determined by solving the corresponding non-relativistic
Schrodinger equation in order to obtain the expected masses of the charmonium
spectrum, characterized by the radial quantum number n and the relative orbital
angular momentum between the quark and the antiquark, L. In particular, all predicted states lying under the 
$\D\Db$
mass threshold have been
observed\cite{Godfrey:2008nc,Swanson:2005tq,Drenska:2010kg,Brambilla:2010cs}. 

On the other hand, the possible existence of more sophisticated states than
mesons and baryons, like the multiquark states, hybrid mesons and mesonic
molecules has been
discussed since the early days of the quark model\cite{Swanson:2005tq,
Close:2008hv, Rosner:2006sv,Ebert:2005nc,Maiani:2005pe}. 

In the last decade, considerable experimental evidence has been collected
about the existence of new states, lying in the charmonium mass range,
but not fitting well the charmonium mass spectrum
picture\cite{:2007wga,Aubert:2005rm, He:2006kg,:2007sj,Abe:2007jn,Abe:2004zs,
Uehara:2005qd}. Most of the observations
also suggested that these candidates are exotic.  These studies have been
performed at Babar and
Belle, two experiments which took data at the \epem Beauty Factories at SLAC
(Stanford Linear Accelerator Center, USA) and KEK (High-Energy Accelerator
Research Organization, Japan), respectively. Confirmations have also come
from the CDF experiment, collecting data from $\proton\antiproton$ interactions at Fermilab,USA.

In these notes  the main results from LHC experiments on XYZ
charmonium-like candidates are summarized. In Section
1 the main features of the \lhcb detector are presented. Section 2 is
dedicated to the discussion of the measurements
of the \Xa mass and cross-section in the \lhcb and \cms experiments. In Section
4, the results of the search 
for the \Xb and \Xc states in \BToKJpsiPhi decays at \lhcb are presented. The
conclusions are presented in Section 5.

\section{LHCb and CMS detectors}

LHCb is an experiment dedicated to heavy flavour physics at the LHC\cite{Alves_2008zz}. 
Its primary goal is to search for indirect evidence of new physics in CP violation 
and rare decays of beauty and charm hadrons.

\begin{figure}[t]
    \centering
    \includegraphics[width=.9\linewidth]{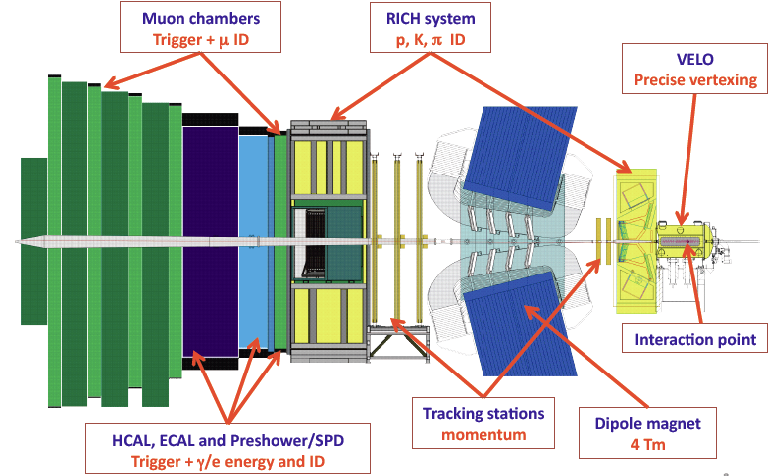}
    \caption{YZ view of the LHCb detector.}
    \label{fig:figure1}
\end{figure}

LHCb detector is a single-arm spectrometer (see figure \ref{fig:figure1}) with a forward angular coverage from approximately 10 mrad
to 300 (250) mrad in the bending (non-bending) plane, corresponding to a pseudorapidity range of 2  $<\eta <$ 5.
In fact, the detector geometry is optimized to cover the region where the \bbbar
cross-section peaks in such way
that, even if just covering about 4\% of the solid angle, the LHCb detects about 40\% of heavy quark
hadrons produced in the proton-proton colisions. 
  
The spectrometer consists of a vertex locator, a warm dipole magnet, a tracking system, two
RICH detectors, a calorimeter system and a muon system. The track momenta are measured to a 
precision of $\delta p/p$ between 0.35\% and 0.5\%. The Ring Imaging Cherenkov
Detector (RICH) 
system provides excellent charged
hadron identification in a momentum range 2-100 GeV/c. The calorimeter system identifies 
high transverse energy hadron, electron and photon candidates and provides information for the trigger. 
The muon system provides information for the trigger and muon identification with an efficiency 
of about 95\% for a misidentification rate of about 1-2 \% for momenta above 10 GeV/c.

The luminosity for the LHCb experiment can be tuned 
by changing the beam focus at its interaction point independently
from the other interaction points, allowing LHCb to maintain the optimal luminosity 
in order not to saturate the trigger or to damage the delicate sub-detectors parts.
In fact, due this capability, LHCb was able to keep its luminosity at the 
constant value of $3.5\times10^{32}\,{\rm cm^{-2} s^{-1}}$ during most of 2011
data taking.

The trigger chain is composed by a first level hardware 
trigger and two levels of software triggers. LHCb uses hadrons,
muons, electrons and photons throughout the trigger chain,
maximizing the trigger efficiency on all heavy quark decays and making the 
experiment sensitive to many different final states.  

In 2010 and 2011, the detector recorded about
$1.1\,{\rm fb^{-1}}$ integrated luminosity in proton-proton collisions
 at $\sqrt{s}=7$ TeV, corresponding to 90\% of the luminosity
delivered by the Large Hadron Collider(LHC) to LHCb.

The CMS is a multi-purpose experiment at LHC, designed with the main goal of search
for new physics phenomena at large transverse momentum scales. CMS cover a
rapidity range up to  $\arrowvert\eta\arrowvert< 2.5$ and since \b-quark production peaks at large
rapidities, \cms is most able to search for charmonium-like candidates produced primary 
in the proton-proton collisions. For a complete description of the \cms detector see \cite{cms_paper}.

\section{\Xa mass and cross-section measurements at \lhc and \cms}

The \Xa resonance was discovered in 2003 by the Belle collaboration in the \BToXaK, \XaToJpsiPiPi decay chain \cite{Choi:2003ue}. Its existence was confirmed by the CDF\cite{Acosta:2003zx}, D$\emptyset$ \cite{Abazov:2004kp} and BaBar\cite{Aubert:2004ns} collaborations.

The \Xa mass is currently known with < 1.0\mevcc precision, the dipion mass spectrum in the decay \XaToJpsiPiPi \cite{Choi:2011fc,Abulencia:2005zc} has been studied and the \Xa quantum numbers have been constrained to be either $J^{PC} = 2^{-+}$ or $1^{++}$ \cite{Abulencia:2006ma} and are still not established. However, despite the cumulated experimental and theoretical effort, the nature of the \Xa remains uncertain. Among the possible interpretations for this state currently discussed in the literature, one can remark the mesonic molecule, the hybrid meson and the tetraquark hypotesis. The conventional charmonium interpretation is not excluded. 

\begin{figure}[t]
 \centering
 \includegraphics[width=.9\linewidth]{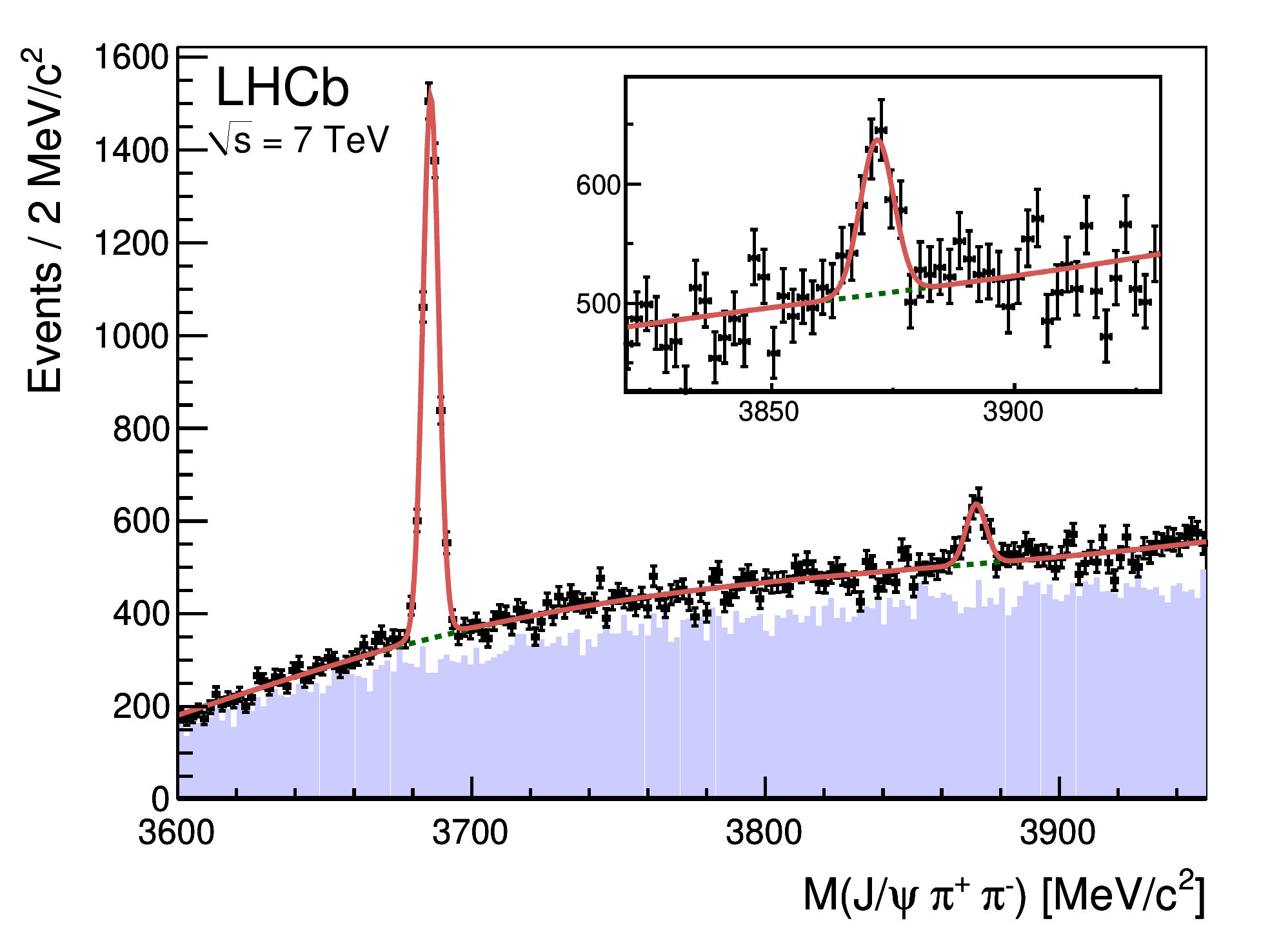}
 \caption{Invariant mass distribution of $\jpsi\pipi$(black points with statistical error bars) and
same-sign $\jpsi\pip\pip$(blue filled histogram) candidates. The solid red curve is the result of the
fit described in the text. The inset shows a zoom of the \Xa region.}
 \label{fig:x-lhcb}
\end{figure}

In  \lhcb the analysis is performed on $34.7\invpb$ dataset collected in 2010 in
$\proton\proton$ collisions at \sqs = 7\tev.
The \Xa signal has been isolated applying tight cuts in order to reduce the combinatorial background, generated when a correctly reconstructed \jpsi meson is combined with a random \pipi pair from the
primary $\proton\proton$ interaction. The selection cuts are optimized
using reconstructed $\psitwos \to \jpsi \pipi$ decays, as well as ``same-sign
pion'' candidates satisfying the same criteria as used for the \Xa and \psitwos
selection. A further background suppression is reached applying the requirement
Q < 300 \mevcc, where $Q =\PM_{\mumu\pipi}-\PM_{\mumu}-\PM_{\pipi}$.
See \cite{Aaij:2011sn} for a detailed discussion on the selection procedure.

The masses of the \psitwos and \Xa mesons are determined from an
extended unbinned maximum likelihood fit of the reconstructed $\jpsi\pipi$
mass in the interval $3.60 < \PM_{\jpsi\pipi} < 3.95 \gevcc$. The \psitwos and
\Xa signals are described with a non-relativistic Breit-Wigner function
convolved with a Gaussian resolution function. The intrinsic width of the
\psitwos is fixed to the PDG value and the \Xa width is fixed to zero in the
nominal fit. The ratio of the mass resolutions for the
\Xa and the \psitwos is fixed to the value $\sigma_{\Xa}/\sigma_{\psitwos} =
1.31$. The background shape is described by the functional form $\Pf(\PM)
\propto (\PM - \PM_{\jpsi}-2\PM_{\pion})^{\Pc_0}\Pe^{(-\Pc_1\PM-\Pc_2\PM^2
)}$.The results of the fit are summarized in the table \ref{tab:fittable}.
\begin{table}
\centering
\begin{tabular}{c|c|c}
  \toprule
  Fit parameter 		& \psitwos  & \Xa \\
  \hline
  Number of signal events	& $3998 \pm 83$ & $565 \pm 62$ \\
  Mass [ \mevcc ]		& $3686.10 \pm 0.06$ & $3871.88 \pm 0.48$\\
  Mass resolution [ \mevcc ]	& $2.54 \pm 0.06$ & $3.33 \pm 0.08$\\
  S/B in $\pm3\sigma$ window		& 1.5 & 0.15\\
  Number of background events	& $73094 \pm 282$ & --\\
  \toprule
\end{tabular}
\caption{Fit results on LHCb \Xa studies.}
\label{tab:fittable}   
\end{table}

At LHCb, the same sample used to measure the \Xa mass has been used to perform 
\Xa production studies. The product of the inclusive production
cross-section $\upsigma(\proton\proton\to\Xa + \cdots)$ and the branching fraction $\BR(\Xa\to\jpsi\pipi)$
is determined according the expression 
\[ \upsigma(\proton\proton\to\Xa
 + \cdots)\times\BR(\Xa\to\jpsi\pipi) = \frac{\PN^{corr}_{\Xa}}{\xi\times\lum_{\rm int} \times \BR(\jpsi\to\mumu) }\]
where $\PN_{\Xa}$ is the efficiency-corrected signal yield, $\upxi$ is a correction factor to the
simulation-derived efficiency that accounts for known differences between data and simulation, $\BR(\jpsi\to\mumu) = (5.93 \pm 0.06)\times10^{-2}$ is the $\jpsi\to\mumu$ branching fraction,
and $\lum_{\rm int}$ is the integrated luminosity. See \cite{Aaij:2011sn} for detailed discussion 
about the calibration procedure and the treatment of the different sources of systematic uncertainty.
The studies are performed just considering candidates lying inside the fiducial region for the measurement defined by 
\[2.5 < y < 4.5 \text{ and } 5 < p_T < 20 \gevc \] where $y$ and $p_T$ are the
rapidity and transverse momentum of the \Xa. 
The X(3872) production cross section at LHCb is measured to be 
 
\[ \upsigma(\proton\proton\to\Xa
 + \cdots)\times\BR(\Xa\to\jpsi\pipi) = 4.7 \pm 1.1\stat \pm 0.7\syst
\nb \]

The \cms Collabration also performed studies on the \Xa production. \cms uses a
dataset of 40\invpb collected in $\proton\proton$ collisions at \sqs=7\tev to
measure the ratio of the branching fractions of \psitwos\to\jpsi\pipi and
\Xa\to\jpsi\pipi which is defined as 
\[ R =  \frac{\sigma({\rm pp}\to\Xa+
\cdots)\times\mathcal{B}(\Xa\to\jpsi\pipi)}{\upsigma({\rm pp}\to\Xa+
\cdots)\times\mathcal{B}(\psitwos \to \jpsi\pipi)}\]
inside the fiducial region defined by  \[\pt > 8\gevc \text{ and } |y| < 2.2\].

 The result of the \cms analysis is
 \[ R =  0.087 \pm 0.017\stat \pm
0.009\syst, \]
where the first error refers to the statistical uncertainty and the
second error contains
the sum of all systematic uncertainties, as described
in \cite{CMS-PAS-BPH}, added in quadrature. See \cite{CMS-PAS-BPH}
for a detailed discussion of the selection procedure and uncertainties
estimation.

\section{Search for the X(4140) state in \BToKJpsiPhi decays}

The CDF collaboration has reported a $3.8\sigma$ evidence for the \XbToJpsiPhi state 
using data collected in proton-antiproton collisions 
at the Tevatron ($\sqrt{s}$ = 1.96 TeV)\cite{PhysRevLett.102.242002}. 
In a preliminary update on the analysis \cite{Aaltonen:2011at}, the CDF collaboration reported $115\pm12$
\BToKJpsiPhi events and $19\pm6$ \Xb candidates with a statistical significance 
of more than $5\sigma$. The mass and width were determined to be
$4143.4^{+2.9}_{-3.0}\pm0.6$MeV/$c^2$ and $15.3^{10.4}_{-6.1}\pm2.5$ MeV/$c^2$, respectively.
The relative branching ratio was measured to be 
$\mathcal{B}(\BToXbK)\times\mathcal{B}(\XbToJpsiPhi)/\mathcal{B}(\BToKJpsiPhi) = 0.149\pm0.039\stat\pm0.024\syst$.

Since a charmonium state at this mass is expected to have much larger width because of
open flavor decay channels, the decay rate of the \XbToJpsiPhi mode, so near
to kinematic threshold, should be small and unobservable. Due to these issues, the CDF's report 
rejuvenated the discussions on exotic hadronic states. It was cogitated that the \Xb
resonance could be a molecular state \cite{Zhi-Gang, Albuquerque2009186,Jian-Rong}, a tetraquark state 
\cite{Stancu, PhysRevD.79.077502}, a hybrid state \cite{Mahajan2009228, springerlink:10.1140} 
or even a rescattering effect [15, 16]. 

The CDF data also suggested the presence of a second state, referred here as \Xc with mass
$4274.4^{+8.4}_{-6.4} \pm 1.9$ MeV/$c^2$ and width  $32.3^{+21.9}_{-15.3}\pm{7.6}$ MeV/$c^2$. 
The corresponding event yield was $22\pm8$ with $3.1\sigma$ significance. 
This observation has also received
attention in the literature \cite{Liu2009137,Finazzo2011101}. 
On the other hand, the Belle experiment found no evidence for the \Xb and \Xc 
states\cite{PhysRevLett.104.112004, Brodzicka}. 

\begin{figure}[t]
    \centering
    \includegraphics[width=.7\linewidth]{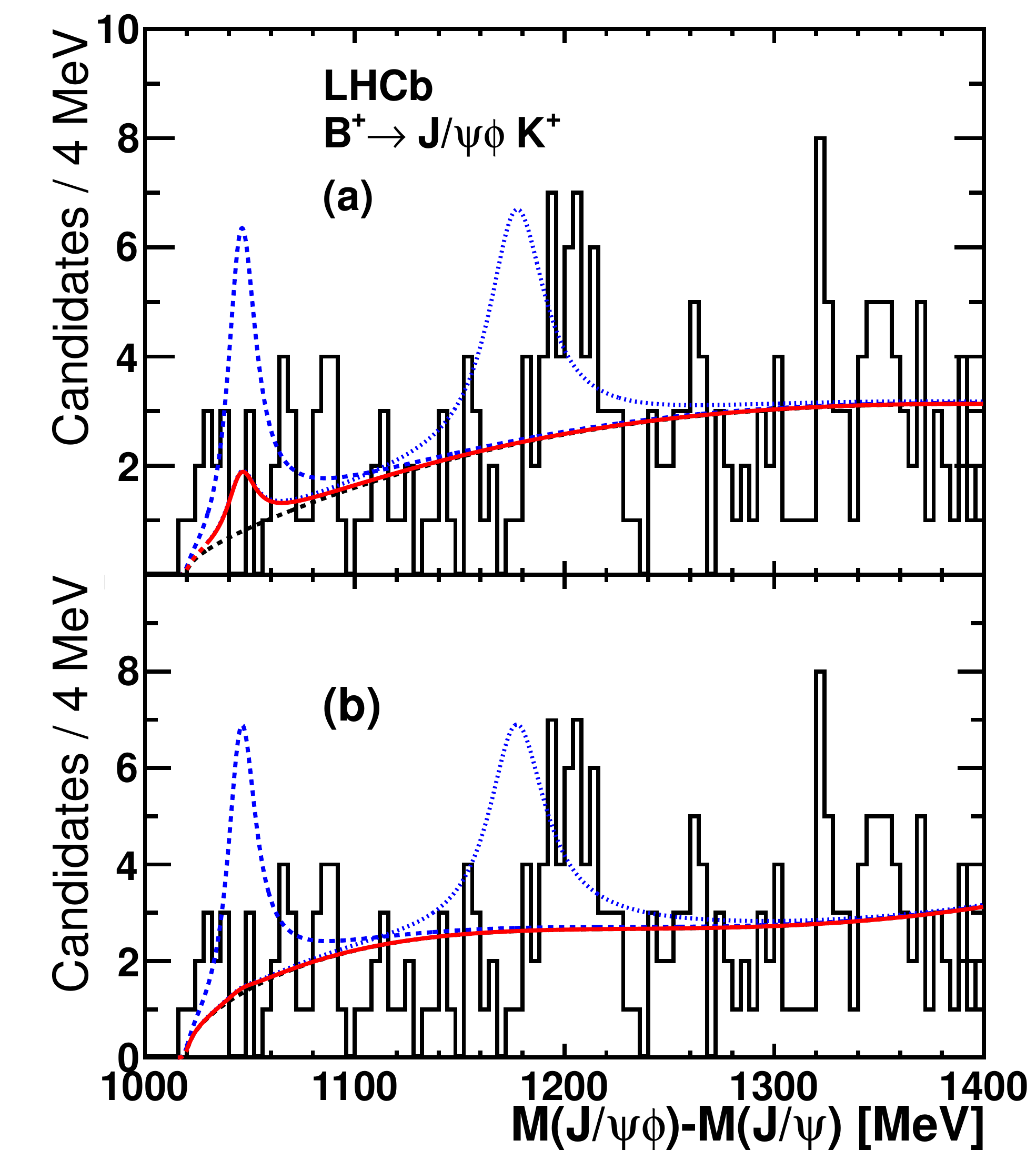}
\caption{Distribution of the mass difference $M(\jpsi\phi)-M(\jpsi)$. Fit of the \Xb signal on top of a smooth
background is superimposed (solid red line). The dashed blue (dotted blue) line on top illustrates
the expected \Xb (\Xc) signal yield from the CDF measurement. The top and
bottom plots differ by the background function (dashed black line) used in the fit: (a) a
background efficiency-corrected three-body phase-space; (b) background efficiency-corrected quadratic function.}
 \label{fig:ECP_figure3}
\end{figure}

The LHCb analysis\cite{Aaij:2012pz, LHCb-CONF-2011-045} starts reconstructing a \Bp candidate as five-track
$(\mu^+\mu^-\Kp\Km\Kp)$ vertex using well reconstructed and identified muons and kaons candidates.
The \Bp  candidates are required to have $p_{T}>4.0$ GeV/$c$ and a decay time of at least 0.25 ps. 
The invariant mass of the $(\mu^+\mu^-\Kp\Km\Kp)$ combination is evaluated after the muon pair 
is constrained to the \jpsi mass, and all final state particles are constrained to a common vertex.
Further background suppression is provided using the likelihood ratio discriminator method.

The \BToKJpsiPhi invariant mass distribution, with at least one \Kp\Km
combination having an invariant mass within $\pm15$ MeV/$c^2$ of the nominal $\phi$ mass was fitted
by a Gaussian and a quadratic function resulting in $346\pm20$ \Bp 
events with a mass resolution of $5.2 \pm 0.3$ MeV/$c^2$. 

The \Xb state was searched selecting events within $\pm15$ MeV/$c^2$ of the $\phi$ mass. 
Figure \ref{fig:ECP_figure3} shows the mass difference $M(\jpsi\phi)-M(\jpsi)$ distribution without 
\jpsi or $\phi$ mass constraints. No narrow structure is observed near the threshold. 
The fit results are $N_{\Xb}^{(a)} = 6.9 \pm 4.9$ or $N_{\Xb}^{(b)} = 0.6 \pm 7.1$,
depending on the background shape used.

The CDF's fit model was used to quantify the compatibility of the two measurements and 
considering the LHCb \BToKJpsiPhi yield, the efficiency ratio, and the CDF value for
$\mathcal{B}(\BToXbK)/\mathcal{B}(\BToKJpsiPhi)$, one concludes that LHCb should have
observed $35 \pm 9 \pm 6$ events, where the first uncertainty is statistical from the CDF data
and the second includes both the CDF and LHCb systematic uncertainties. 
The LHCb results disagree with the CDF observation by $\sim 2.7\sigma$.
In the case of the \Xc candidate, the same procedure predicts that LHCb should
have observed $53\pm19$ \Xc candidates.
The final results are the following upper limits at 90\%CL

\[\frac{\mathcal{B}(\BToXbK)\times\mathcal{B}(\XbToJpsiPhi)}{\mathcal{B}
(\BToKJpsiPhi)} < 0.07,\]
\[\frac{\mathcal{B}(\BToXcK)\times\mathcal{B}(\XcToJpsiPhi)}{\mathcal{B}
(\BToKJpsiPhi)} < 0.08.\]


\section{Conclusions}

A selection of results on XYZ states spectroscopy at the LHC have
been summarized.
Many new results are expected from the analysis 
of the 2011 and 2012 datasets and 
as well from the news studies currently on-going.

The LHC experiments are in a privileged position to explore the production
mechanisms and spectra of the XYZ states, delivering competitive
 results in the heavy flavor sector.

\bibliographystyle{h-physrev.bst}
\bibliography{proceedings}

\begin{thebibliography}{10}

\bibitem{Godfrey:2008nc}
S.~Godfrey and S.~L. Olsen,
\newblock Ann.Rev.Nucl.Part.Sci. {\bf 58}, 51 (2008), 0801.3867.

\bibitem{Swanson:2005tq}
E.~Swanson,
\newblock Int. J. Mod. Phys. {\bf A21}, 733 (2006), hep-ph/0509327.

\bibitem{Drenska:2010kg}
N.~Drenska {\em et~al.},
\newblock Riv. Nuovo Cim. {\bf 033}, 633 (2010), 1006.2741.

\bibitem{Brambilla:2010cs}
N.~Brambilla {\em et~al.},
\newblock Eur.Phys.J. {\bf C71}, 1534 (2011), 1010.5827.

\bibitem{Close:2008hv}
F.~E. Close,
\newblock (2008), 0801.2646.

\bibitem{Rosner:2006sv}
J.~L. Rosner,
\newblock J. Phys. Conf. Ser. {\bf 69}, 012002 (2007), hep-ph/0612332.

\bibitem{Ebert:2005nc}
D.~Ebert, R.~N. Faustov, and V.~O. Galkin,
\newblock Phys. Lett. {\bf B634}, 214 (2006), hep-ph/0512230.

\bibitem{Maiani:2005pe}
L.~Maiani, V.~Riquer, F.~Piccinini, and A.~D. Polosa,
\newblock Phys. Rev. {\bf D72}, 031502 (2005), hep-ph/0507062.

\bibitem{:2007wga}
S.~K. Choi {\em et~al.}, BELLE,
\newblock Phys. Rev. Lett. {\bf 100}, 142001 (2008), 0708.1790.

\bibitem{Aubert:2005rm}
B.~Aubert {\em et~al.}, BABAR,
\newblock Phys. Rev. Lett. {\bf 95}, 142001 (2005), hep-ex/0506081.

\bibitem{He:2006kg}
Q.~He {\em et~al.}, CLEO,
\newblock Phys. Rev. {\bf D74}, 091104 (2006), hep-ex/0611021.

\bibitem{:2007sj}
C.~Z. Yuan {\em et~al.}, Belle,
\newblock Phys. Rev. Lett. {\bf 99}, 182004 (2007), 0707.2541.

\bibitem{Abe:2007jn}
K.~Abe {\em et~al.}, Belle,
\newblock Phys. Rev. Lett. {\bf 98}, 082001 (2007), hep-ex/0507019.

\bibitem{Abe:2004zs}
K.~Abe {\em et~al.}, Belle,
\newblock Phys. Rev. Lett. {\bf 94}, 182002 (2005), hep-ex/0408126.

\bibitem{Uehara:2005qd}
S.~Uehara {\em et~al.}, Belle,
\newblock Phys. Rev. Lett. {\bf 96}, 082003 (2006), hep-ex/0512035.

\bibitem{Alves_2008zz}
LHCb Collaboration,
\newblock JINST {\bf 3}, S08005 (2008).

\bibitem{cms_paper}
S.~Chatrchyan {\em et~al.}, CMS Collaboration,
\newblock Journal of Instrumentation {\bf 3}, S08004 (2008).

\bibitem{Choi:2003ue}
S.~Choi {\em et~al.}, Belle Collaboration,
\newblock Phys.Rev.Lett. {\bf 91}, 262001 (2003), hep-ex/0309032.

\bibitem{Acosta:2003zx}
D.~Acosta {\em et~al.}, CDF Collaboration,
\newblock Phys.Rev.Lett. {\bf 93}, 072001 (2004), hep-ex/0312021.

\bibitem{Abazov:2004kp}
V.~Abazov {\em et~al.}, D0 Collaboration,
\newblock Phys.Rev.Lett. {\bf 93}, 162002 (2004), hep-ex/0405004.

\bibitem{Aubert:2004ns}
B.~Aubert {\em et~al.}, BABAR Collaboration,
\newblock Phys.Rev. {\bf D71}, 071103 (2005), hep-ex/0406022.

\bibitem{Choi:2011fc}
S.-K. Choi {\em et~al.},
\newblock Phys.Rev. {\bf D84}, 052004 (2011), 1107.0163.

\bibitem{Abulencia:2005zc}
A.~Abulencia {\em et~al.}, CDF Collaboration,
\newblock Phys.Rev.Lett. {\bf 96}, 102002 (2006), hep-ex/0512074.

\bibitem{Abulencia:2006ma}
A.~Abulencia {\em et~al.}, CDF Collaboration,
\newblock Phys.Rev.Lett. {\bf 98}, 132002 (2007), hep-ex/0612053.

\bibitem{Aaij:2011sn}
LHCb Collaboration,
\newblock hep-ex/1112.5310.

\bibitem{CMS-PAS-BPH}
S.~Chatrchyan {\em et~al.}, CMS Collaboration,
\newblock CMS-PAS-BPH-10-018  (2011).

\bibitem{PhysRevLett.102.242002}
CDF Collaboration,
\newblock Phys. Rev. Lett. {\bf 102}, 242002 (2009).

\bibitem{Aaltonen:2011at}
CDF Collaboration,
\newblock (2011), hep-ex/1101.6058.

\bibitem{Zhi-Gang}
Z.-G. Wang, Z.-C. Liu, and X.-H. Zhang,
\newblock The European Physical Journal C - Particles and Fields {\bf 64}, 373
  (2009),
\newblock 10.1140/epjc/s10052-009-1156-2.

\bibitem{Albuquerque2009186}
R.~M. Albuquerque, M.~E. Bracco, and M.~Nielsen,
\newblock Physics Letters B {\bf 678}, 186 (2009).

\bibitem{Jian-Rong}
J.-R. Zhang and M.-Q. Huang,
\newblock Journal of Physics G: Nuclear and Particle Physics {\bf 37}, 025005
  (2010).

\bibitem{Stancu}
F.~Stancu,
\newblock Journal of Physics G: Nuclear and Particle Physics {\bf 37}, 075017
  (2010).

\bibitem{PhysRevD.79.077502}
N.~V. Drenska, R.~Faccini, and A.~D. Polosa,
\newblock Phys. Rev. D {\bf 79}, 077502 (2009).

\bibitem{Mahajan2009228}
N.~Mahajan,
\newblock Physics Letters B {\bf 679}, 228 (2009).

\bibitem{springerlink:10.1140}
Z.-G. Wang,
\newblock The European Physical Journal C - Particles and Fields {\bf 63}, 115
  (2009),
\newblock 10.1140/epjc/s10052-009-1097-9.

\bibitem{Liu2009137}
X.~Liu,
\newblock Physics Letters B {\bf 680}, 137 (2009).

\bibitem{Finazzo2011101}
S.~I. Finazzo, M.~Nielsen, and X.~Liu,
\newblock Physics Letters B {\bf 701}, 101 (2011).

\bibitem{PhysRevLett.104.112004}
C.~P. Shen {\em et~al.}, Belle Collaboration,
\newblock Phys. Rev. Lett. {\bf 104}, 112004 (2010).

\bibitem{Brodzicka}
{J. Brodzicka},
\newblock Heavy flavour spectroscopy,
\newblock in {\em Lepton Photon 2009 (LP09)} Vol. DESY-PROC-2010-04, 2010.

\bibitem{Aaij:2012pz}
LHCb Collaboration,
\newblock hep-ex/1202.5087.

\bibitem{LHCb-CONF-2011-045}
LHCb Collaboration,
\newblock LHCb-CONF-2011-045,.

\end{thebibliography}
\end{document}